\documentclass[aps,prb,reprint]{revtex4-2}
\usepackage[dvipdfmx]{graphicx}
\usepackage{amsmath,amssymb}
\usepackage{color}
\usepackage{bm}
\usepackage{hyperref}
\usepackage{comment}
\usepackage[normalem]{ulem}

\begin{document}

\title{Minimal alternating current injection into Carbon Nanotubes}

\author{Kota Fukuzawa$^{1}$, Takeo Kato$^{1}$, Thibaut Jonckheere$^{2}$, J\'{e}r\^{o}me Rech$^{2}$, and Thierry Martin$^{2}$}
\affiliation{
${^1}$Institute for Solid State Physics, The University of Tokyo, Kashiwa, 277-8581, Japan\\
${^2}$Aix Marseille Univ, Universit\'e de Toulon, CNRS, CPT, Marseille, France
}

\date{\today}

\begin{abstract}
We study theoretically the effect of electronic interactions in 1d systems on electron injection using periodic Lorentzian pulses, known as Levitons.
We consider specifically a system composed of a metallic single-wall carbon nanotube, 
described with the Luttinger liquid formalism, a scanning tunneling microscope (STM) tip, and metallic leads.  
Using the out-of-equilibrium Keldysh Green function formalism,
we compute the
current and current noise in the system.
We prove that the excess noise vanishes when each Leviton injects an integer number of electrons from the STM tip into the nanotube.
This extends the concept of minimal injection with Levitons to
strongly correlated, uni-dimensional non-chiral systems.
We also study the time-dependent current profile, and show how it is the result of  interferences between pulses non-trivially reflected at the nanotube-lead interface.
\end{abstract}

\maketitle 

\section{Introduction}
\label{sec:introduction}

Controlled electron injection in an electronic system is an important issue, both for potential applications to electron quantum optics and for the fundamental study of the many-body properties of the system~\cite{feve07, hermelin11, mcneil11,Charles2011,bocquillon12,Bocquillon2014,Bauerle2018}.
More than twenty years ago, Levitov and co-workers have shown that applying a generic time-dependent voltage to inject a charge creates a fundamental disturbance to the system, akin to the Anderson catastrophe, with the creation of a divergent number of electron-hole pairs in the Fermi sea, in addition to the injected charge~\cite{levitov96,ivanov97,keeling06}.
Importantly, they also showed that, by applying a specific quantized time-dependent voltage on a electronic conductor, it is possible to excite a single electron above the Fermi sea, without creating any perturbation to the system.
These peculiar excitations have been called \textit{Levitons}, and have been realized experimentally in 2d electron gases~\cite{dubois13,Bocquillon2014,jullien14}.
Such injection of a single charge, without any spurious excitation of the system, has been called \textit{minimal injection}, and can be characterized by studying the excess noise.
The properties of the Levitons have been studied intensively in various systems~\cite{Battista2014,Forrester2014,Dasenbrook2015,Dasenbrook2016,Moskalets2016,Suzuki2017,Cabart2018,Moskalets2018,Nastaran2019,Bisognin2019,Burset2019}.
   
Coulomb interactions can have a major impact on the many-body state of electronic systems. 
This is particularly true for 1d electronic systems, where interactions lead to a very specific behavior, which can be described at low excitation energies by the Luttinger liquid theory~\cite{giamarchi03}.
This applies for example for conducting carbon nanotubes~\cite{Egger1998,kane97,Yoshioka1999}, semiconductor nanowires~\cite{zaitsev00,Auslaender2002}, edge states of the fractional quantum Hall effect~\cite{wen95,Chang2003}, etc.
Important physical quantities of these systems, such as the tunneling density of states and the current voltage characteristics typically obey some power-law behavior, with an exponent which is explicitly dependent on the interaction parameter.
Other remarkable behaviors of 1d interacting systems are the charge fractionalization thanks to the existence of collective mode~\cite{Pham2000,Imura2002,Steinberg2008}, spin-charge separation~\cite{Jompol2009}, etc.
When connected to standard, non-interacting electrodes, non trivial processes at the boundaries between the interacting system and the electrodes, similar to Andreev reflection, do occur, creating a complex behavior for the time-dependent current~\cite{maslov95,safi95,ponomarenko95,sandler98,Acciai2019,hashisaka21,glidic2023}. 

\begin{figure}[tbp]
  \begin{center}
    \includegraphics[width=7cm]{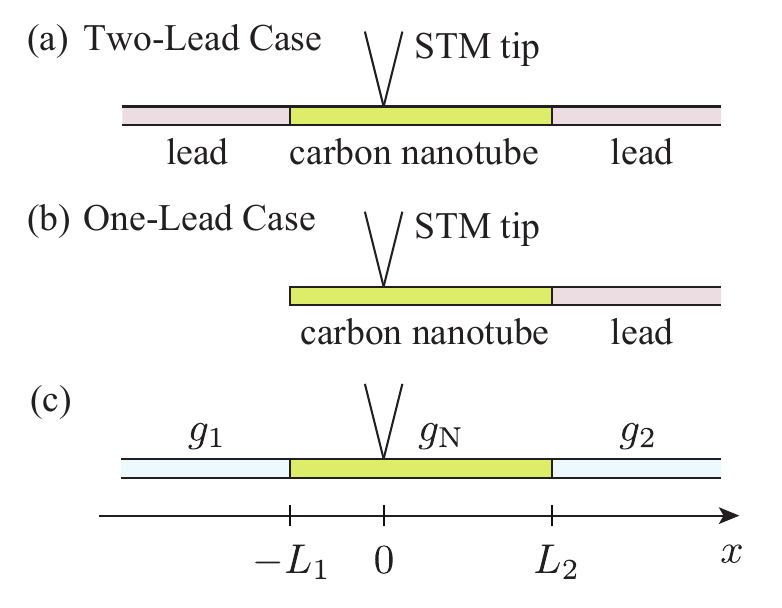}
    \caption{Schematic of the system composed of a carbon nanotube, metallic leads, and a STM tip. (a) Two-lead setup. (b) One-lead setup. (c) A model for general theoretical description.}
    \label{fig:Setup}
  \end{center}
\end{figure}

It is thus natural to ask whether the concept of \textit{Levitons} can be extended to 1d interacting electronic systems.
This question has already been given a positive answer in the case of chiral edge states of FQHE~\cite{rech17}.
However, in a non-chiral system, the behavior at the interface between the interacting system and a normal lead is much more complex, which may modify strongly the physical response of the system to an AC charge injection.
To answer this question in the case of a non-chiral system, in this work we consider explicitly AC electron injection into a carbon nanotube (CNT) from a STM tip~\cite{Crepieux2003,Lebedev2005,Guigou2007,Guigou2009a,Guigou2009b} (see Fig.~\ref{fig:Setup}), and calculate the excess noise induced by the external AC driving.
While the STM-CNT coupling is treated perturbatively as it operates in the weak tunneling regime, we treat the Coulomb interactions and the coupling between the CNT and the leads non-perturbatively using the bosonization method.  
We show that, although interactions have a deep impact on the behavior of the system, it is still possible to inject a single electronic charge with zero excess noise by using Leviton pulses.
Furthermore, we provide an explicit general proof that the excess noise is zero for Levitons which inject an integer number of electrons.
We also discuss the formulas for the noise and current, which contains the complex physics of Andreev-like reflection at the interface with the leads.

The rest of this work is organized as follows. 
In Sec.~\ref{sec:system}, we briefly summarize our theoretical model for the CNT, metallic leads, and STM tip.
We also formulate the external voltage driving through the STM-CNT coupling.
In Sec.~\ref{sec:formulation}, we show the analytical formulas for the tunneling current, noise and excess noise (details of the derivation of these formulas are given in the two Appendices).
In Sec.~\ref{sec:ExcessNoise}, we show the numerical results for the excess noise for different values of the relevant parameters, and discuss its properties.
We also provide an analytic proof that the excess noise is zero for Levitons with integer charge, thus answering rigorously the initial question motivating this work.
Finally, to clarify the importance of the Andreev-like reflection at the CNT-lead junction, we discuss the current profile as a function of time in Sec.~\ref{sec:CurrentProfiles}.
We summarize our results in Sec.~\ref{sec:Summary}.

\section{Description of the system}
\label{sec:system}

We consider a metallic CNT connected at both ends to two semi-infinite Fermi liquid leads (see Fig.~\ref{fig:Setup}~(a)) or on one end to one Fermi liquid semi-infinite lead (see Fig.~\ref{fig:Setup}~(b)).
A STM tip is placed close to the CNT, and a time-dependent voltage $V(t)$ is applied between the tip and the CNT to allow the tunneling of electrons.
These two setups can be described generally by an infinite one-dimensional system with inhomogeneous interaction parameters as shown in Fig.~\ref{fig:Setup}~(c), whose Hamiltonian is given with the bosonization technique as~\cite{Crepieux2003,Lebedev2005,Guigou2007}
\begin{align}
\hat{H}=\sum_{j\delta}\int  dx \left[ \frac{v_{j\delta}(x)g_{j\delta}(x)}{2}(\partial_x \phi_{j\delta})^2+\frac{v_{j\delta}(x)}{2g_{j\delta}(x)}(\partial_x \theta_{j\delta})^2 \right].
\label{HamCNT}
\end{align}
Here, $\phi_{j\delta}(x)$ and $\theta_{j \delta}(x)$ are non-chiral bosonic fields which satisfy the commutation relation 
\begin{align}
    [\phi_{j\delta}(x),\theta_{j'\delta'}(x')]=-(i/2)\delta_{jj'}\delta_{\delta\delta'} \mathrm{sgn}(x-x'),
\end{align}
$g_{j \delta}(x)$ is the interaction parameter, 
$v_{j \delta}(x)=v_{\rm F}/g_{j \delta}(x)$ is the renormalized velocity, $v_{\rm F}$ is the Fermi velocity in the absence of Coulomb interaction,
$j \in ({\rm c},{\rm s})$ specifies charge and spin sectors, and $\delta \in (+,-)$ specifies the symmetric and anti-symmetric sectors with respect to the branch (the valley) in the CNT.
The origin of the coordinate $x$ is set to the position of the STM tip.
Assuming that the screened Coulomb interaction modifies only the $(j,\delta) = (c,+)$ sector~\cite{Egger1998,Yoshioka1999},
the interaction parameters are set as $g_{c-} = g_{s+} = g_{s-}= 1$ and
\begin{align}
g_{c+} = \left\{ \begin{array}{ll}
g_1, & (x<-L_1), \\
g_{\rm N}, & (-L_1 < x < L_2), \\
g_2, & (L_2<x),
\end{array} \right. 
\end{align}
where $L_1$ and $L_2$ are lengths of the CNT separated by the STM tip (see Fig.~\ref{fig:Setup}~(c)).
We also define the total length of the CNT as $L=L_1 + L_2$.
The two-lead case shown in Fig.~\ref{fig:Setup}~(a) can be described by setting $g_1 = g_2 = 1$.
On the other hand, the one-lead case shown in Fig.~\ref{fig:Setup}~(b) can be described by setting $g_1 = 0$ to express an open boundary at $x=-L_1$ while $g_2$ ($=1$) is left unchanged.
We note that for repulsive Coulomb interaction $g_{\rm N}$ becomes smaller than unity. 

The electron operator in the CNT is given by
\begin{align}
\Psi_{r\alpha\sigma}(x)=\frac{\eta_{r\alpha \sigma}}{\sqrt{2\pi a}}e^{iq_{\rm F} rx+ik_{\rm F} \alpha x+i\varphi_{r\alpha\sigma}(x)} ,
\end{align}
where $\eta_{r\alpha \sigma}$ is a Klein factor, $a$ is a short-length cutoff, $k_{\rm F}$ is the Fermi wavenumber, and $q_{\rm F}$ ($\ll k_{\rm F}$) is the momentum mismatch associated with the two modes.
Hereafter, we neglect the Klein factor as it does not affect the results. 
The bosonic field $\varphi_{r\alpha \sigma}$ is described with $\phi_{j\delta}(x)$ and $\theta_{j \delta}(x)$ as
\begin{align}
\varphi_{r\alpha \sigma}(x)=\frac{\sqrt{\pi}}{2}\sum_{j\delta}h_{\alpha \sigma j \delta}[\phi_{j\delta}(x)+r\theta_{j\delta}(x)],
\label{eq:varphi}
\end{align}
where $h_{\alpha \sigma c+}=1$, $h_{\alpha \sigma c-}=\alpha$, $h_{\alpha \sigma s+}=\sigma$, and $h_{\alpha \sigma s-}=\alpha\sigma$.
For convenience, the STM is also modeled as a one-dimensional non-interacting system in a bosonized form
\begin{align}
c_{\sigma}(t)=\frac{1}{\sqrt{2\pi a}}e^{i\Tilde{\varphi}_\sigma (t)}.
\end{align}

The electron tunneling between the STM tip and the CNT is described by the Hamiltonian
\begin{align}
\hat{H}_T(t) &= \sum_{r\alpha \sigma \epsilon}\epsilon \Gamma_\epsilon(t)
\Psi_{r\alpha \sigma}^{(-\epsilon)}(0,t) c_\sigma^{(\epsilon)}(t),
\label{eq:HT} \\
\Gamma_\epsilon(t) &= \Gamma \exp \left[ \frac{i\epsilon e}{\hbar} \int_{-\infty}^{t}V(t')\, dt'\right],
\end{align}
where $\Gamma$ and $\Gamma_\epsilon(t)$ are the tunneling amplitudes without and with effect of the time-dependent voltage $V(t)$ and the superscript $\epsilon$ leaves either
operator unchanged ($\epsilon= +$) or transforms it into its Hermitian
conjugate ($\epsilon= -$).
The applied voltage $V(t)$ is divided into DC and AC parts as $V(t) = V_{\rm dc} + V_{\rm ac}(t)$, where by definition $V_{\rm ac}(t)$ averages to zero over one drive period $T$.
Here, we consider three types of the voltage pulse:
\begin{align}
& {\rm Leviton: \ } V(t) = \frac{V_{\rm dc}}{\pi}\sum_k \frac{\eta}{\eta^2+(t/T-k)^2}, 
\label{eq:PulseShape1}\\
& {\rm Cosine: \ } V(t) = 
V_{\rm dc}(1-\cos \Omega t), 
\label{eq:PulseShape2}\\
& {\rm Square: \ } V(t) = 2 V_{\rm dc}\sum_k {\rm rect}(2t/T-k), 
\label{eq:PulseShape3}
\end{align}
where $\Omega$ is the driving frequency, $T=2\pi/\Omega$ is the period, ${\rm rect}(x)=1$ for $x<1/2$
($=0$, otherwise) is the rectangular function, and $\eta = W/T$ ($W$: the half-width at half-maximum of the Lorentzian pulse).
We define the Fourier components of $\Gamma_\epsilon(t)$ as
\begin{align}
\Gamma_-(t) = \Gamma \sum_{l=-\infty}^\infty p_l e^{-i(\omega_0 + l\Omega)t}, 
\end{align}
and $\Gamma_+(t) = (\Gamma_-(t))^*$, where $\omega_0 \equiv eV_{\rm dc}/\hbar$ and 
\begin{align}
p_l & \equiv \int_{-T/2}^{T/2} \frac{dt}{T} e^{il\Omega t}\exp\left[-\frac{ie}{\hbar} \int_{-\infty}^t dt' V_\text{ac}(t')\right] .
\end{align}
For the pulse shapes given in Eqs.~(\ref{eq:PulseShape1})-(\ref{eq:PulseShape3}), the Fourier components are given by
\begin{align}
& {\rm Leviton:\ } p_l = 
q\sum_{s=0}^\infty \frac{\Gamma(q+l+s)}{\Gamma(q+1-s)} \frac{(-1)^s e^{-2\pi\eta(2s+l)}}{(l+s)!s!}, \\
& {\rm Cosine: \ } p_l = J_l(-q), \\
& {\rm Square: \ } p_l = \frac{2}{\pi} \frac{q}{l^2-q^2} \sin \left[ \frac{\pi}{2} (l-q) \right], 
\end{align}
where $J_l(z)$ is the Bessel function.
Here, we introduced a dimensionless  quantity $q$ defined as
\begin{align}
q &\equiv \frac{eV_{\rm dc}}{\hbar \Omega} = \frac{\omega_0}{\Omega} .
\end{align}
This quantity means that the charge injected per period is $q e$. 

\section{Formulation of the excess noise}
\label{sec:formulation}

The current operator is expressed by the bosonic field as~ \cite{Crepieux2003}
\begin{align}
\hat{I}(x,t)=2ev_{\rm F}\frac{\partial_x \phi_{c+}(x,t)}{\sqrt{\pi}}.
\label{eq:CurrentDef}
\end{align}
In the Keldysh formalism, the average current is written in the form:
\begin{align}
I(x,t)&=\frac12 \sum_{\eta} \left \langle T_K{\hat{I}(x,t^\eta)e^{i\int_K H_T(t_1)dt_1}} \right \rangle \nonumber \\
&= \left \langle T_K{\hat{I}(x,t^-)e^{i\int_K H_T(t_1)dt_1}} \right \rangle,
\end{align}
where $T_K$ indicates time-ordering operator along the Keldysh contour, $t^\eta$ indicates time on the forward ($\eta=+$) and backward ($\eta=-$) contour, and $\int_K$ indicates an integral over the Keldysh contour K. 
In the second line, we fixed the time on the backward contour because the current average is independent of $\eta$.
The second-order perturbation with respect to $H_T$ gives
\begin{align}
I(x,t)&=-\frac{1}{2}\sum_{\eta_1\eta_2}\eta_1 \eta_2 \int dt_1 dt_2 \nonumber \\
&\times \langle T_K\{\hat{I}(x,t^-)H_T(t_1^{\eta_1}) H_T(t_2^{\eta_2})\} \rangle .
\end{align}
In the Keldysh formalism, the current fluctuations can be written in the form:
\begin{align}
& S(x,t,t')=\langle \hat{I}(x,t) \hat{I}(x,t')\rangle \nonumber \\
& =\left \langle T_K{\hat{I}(x,t^-)\hat{I}(x,t^+)e^{i\int_K H_T(t_1)dt_1}} \right \rangle.
\end{align}
The second-order perturbation gives
\begin{align}
& S(x,t,t')=-\frac{1}{4}\sum_{\eta \eta_1\eta_2}\eta_1 \eta_2 \int dt_1 dt_2 \nonumber \\ 
& \hspace{10mm} \times \langle T_K\{\hat{I}(x,t^\eta)  \hat{I}(x,t^{-\eta})H_T(t_1^{\eta_1}) H_T(t_2^{\eta_2})\} \rangle .
\label{eq:2ndS}
\end{align}
The current noise $S(x)$ at the position $x$ is obtained by
\begin{align}
S(x) &= \lim_{T\rightarrow 0} \frac{1}{T} \int_{-T/2}^{T/2} dt \int_{-T/2}^{T/2} dt' \, S(x,t,t') \nonumber \\
&= \int_{-\infty}^{\infty} dt \, S(x,t,0).
\end{align}
The excess noise, which is a noise induced by ac driving, is defined as
\begin{align}
S^{\rm ex}(x) = S(x) - e\bar{I}(x) ,
\end{align}
where $\bar{I}(x)$ is the time-averaged current and the second term represents the Poisson noise due to nonequilibrium currents injected from the STM tip.
Hereafter, we focus on the average current and excess noise in the right lead ($x>L_2$).

The results of the second-order perturbation are obtained after a somewhat lengthy 
 calculation, which goes beyond the calculations found in Ref.~\onlinecite{Guigou2007}. As we are interested in the low
temperature behavior of the system, we peform the calculations
at zero temperature only.
We only show the final results for $x>L_2$ as follows:
\begin{widetext}
\begin{align}
\bar{I}(x)
&=\frac{8e \Gamma^2}{N_{\rm lead}\pi^2 a^2} 
\sum_{l=-\infty}^{\infty} \left| p_{l} \right|^2
\int_0^\infty \!\!\!dt \; \sin((q+l) \Omega t) \frac{D(t)}{(1+(v_Ft/a)^2)^{\frac{1}{2}}} \sin(F(t)+\mathrm{ArcTan}(v_Ft/a))
\label{eq:Ix}, \\
S(x)
&=\frac{8e^2 \Gamma^2}{N_{\rm lead} \pi^2 a^2} 
\sum_{l=-\infty}^{\infty} \left| p_{l} \right|^2
\int_0^\infty \!\!\!dt \; \cos((q+l) \Omega t) \frac{D(t)}{(1+(v_Ft/a)^2)^{\frac{1}{2}}} \cos(F(t)+\mathrm{ArcTan}(v_Ft/a)) ,
\label{eq:Sx} \\
S^{\rm ex}(x)
&=\frac{8e^2 \Gamma^2}{N_{\rm lead}\pi^2 a^2} 
\sum_{l=-\infty}^{\infty} \left| p_{l} \right|^2
\int_0^\infty \!\!\!dt \, \frac{D(t)}{(1+(v_Ft/a)^2)^{\frac{1}{2}}} \cos((q+l) \Omega t+ F(t)+\mathrm{ArcTan}(v_Ft/a)) ,
\label{eq:Sexx} 
\end{align}
\begin{align}
F(t)&=\nu \mathrm{ArcTan}\left[\frac{v_Ft}{a}\right] 
+\sum_{k=1}^\infty \frac{(b_1 b_2)^{k}}{8} \left(g_{\rm N}+\frac{1}{g_{\rm N}}\right)\mathrm{ArcTan}\left[\frac{2a v_Ft}{a^2+(2kLg_{\rm N})^2-(v_Ft)^2}\right] \nonumber\\
&+\sum_{k=0}^\infty   \frac{b_1^{k+1}b_2^k}{16} \left(-g_{\rm N}+\frac{1}{g_{\rm N}}\right)\mathrm{ArcTan}\left[\frac{2a v_Ft}{a^2+((2kL+2L_1)g_{\rm N})^2-(v_Ft)^2}\right] \nonumber\\
&\left. +\sum_{k=0}^\infty   \frac{b_1^k b_2^{k+1}}{16} \left(-g_{\rm N}+\frac{1}{g_{\rm N}}\right)\mathrm{ArcTan}\left[\frac{2a v_Ft}{a^2+((2kL+2L_2)g_{\rm N})^2-(v_Ft)^2}\right] \right] ,
\label{eq:Ft}
\end{align}
\begin{align}
D(t)&=\left( \frac{a^2+(v_Ft)^2}{a^2} \right)^{-\nu/2} \nonumber\\
&\times \prod_{k=1}^\infty \left[ \left( \frac{a^2+(2kLg_{\rm N})^2-(v_Ft)^2}{a^2+(2kLg_{\rm N})^2} \right)^2 + \left( \frac{2a v_Ft}{a^2+(2kLg_{\rm N})^2} \right)^2 \right]^{-\frac{(b_1b_2)^{k}}{16}\left( g_{\rm N}+\frac{1}{g_{\rm N}} \right)} \nonumber\\
&\times \prod_{k=0}^\infty \left[ \left( \frac{\alpha^2+((2kL+2L_1)g_{\rm N})^2-(v_Ft)^2}{\alpha^2+((2kL+2L_1)g_{\rm N})^2} \right)^2 + \left( \frac{2\alpha v_Ft}{a^2+((2kL+2L_1)g_{\rm N})^2} \right)^2 \right]^{-\frac{b_1^{k+1}b_2^k}{32}\left(-g_{\rm N}+\frac{1}{g_{\rm N}} \right)} \nonumber\\
&\times \prod_{k=0}^\infty \left[ \left( \frac{a^2+((2kL+2L_2)g_{\rm N})^2-(v_Ft)^2}{a^2+((2kL+2L_2)g_{\rm N})^2} \right)^2 + \left( \frac{2a v_Ft}{a^2+((2kL+2L_2)g_{\rm N})^2} \right)^2 \right]^{-\frac{b_1^k b_2^{k+1}}{32}\left(-g_{\rm N}+\frac{1}{g_{\rm N}} \right)},
\label{eq:Dt}
\end{align}
\end{widetext}
where $\nu=(6+g_{\rm N}+g_{\rm N}^{-1})/8$, $N_{\rm lead}$ is the number of metallic leads and
$b_{1,2} = (g_N - g_{1,2})/(g_N+ g_{1,2})$ are the reflection coefficients. 
We note that the calculated current and current noise are independent of $x$ as far as $x>L_2$. 
Details of the derivation are given in Appendix~\ref{app:Green}.
We also note that the current, the noise and the excess noise include a prefactor $1/N_{\rm lead}$ which reflects the fact that the total injected current is partitioned into the two leads in the two-lead case, while it flows fully into the single lead in the one-lead case. The current and noise are
thus roughly two times larger in the one-lead case compared to the two-lead one (see also Fig.~\ref{fig:2leads} and \ref{fig:1lead} in the next section).

While the formulas for the $F$ and $D$ factors are somewhat heavy,
the physical meaning of the different terms is quite simple. For each term in the sum
defining $F$, there is a corresponding term in the product defining $D$. 
The first term correspond to the contribution for an infinite nanotube,
with direct propagation from the injection point located at the origin to the measurement point $x$. All the other terms correspond to the propagation after a given number
of reflection on the CNT-leads interface: $b_1$ (resp. $b_2$) is the reflection coefficient at the left (resp. right) interface, so for example a 
term with the $(b_1 b_2)^k$ corresponds to $k$ reflection at both interface,
and one can see that this term also contains a factor $2 k L g_N$ which stands for 
$k$ times the length of one round trip along the CNT. Note that $b_1<0$ and $b_2<0$
for the two leads case, which means that the reflections
are similar to an Andreev reflection, with an electron-like excitation
being reflected as a hole-like excitation.

\section{Excess Noise}
\label{sec:ExcessNoise}

In this section, we discuss the properties of the excess noise as a function of $q= e V_{dc}/\hbar \Omega$.
We mainly focus on the excess noise when Leviton pulses are injected.
We define the charge velocity in the CNT as $v_{\rm N}=v_{\rm F}/g_{\rm N}$ and the unit of  time and frequency as $T_0=2L_2/v_{\rm N}$ (the time
needed for an electron to travel from the tip to the right lead and back) and $\Omega_0=2\pi/T_0$, respectively.

\subsection{Two-lead case}
\label{sec:TwoLeadCase}

\begin{figure*}
    \centering
    \includegraphics[width=18.cm]{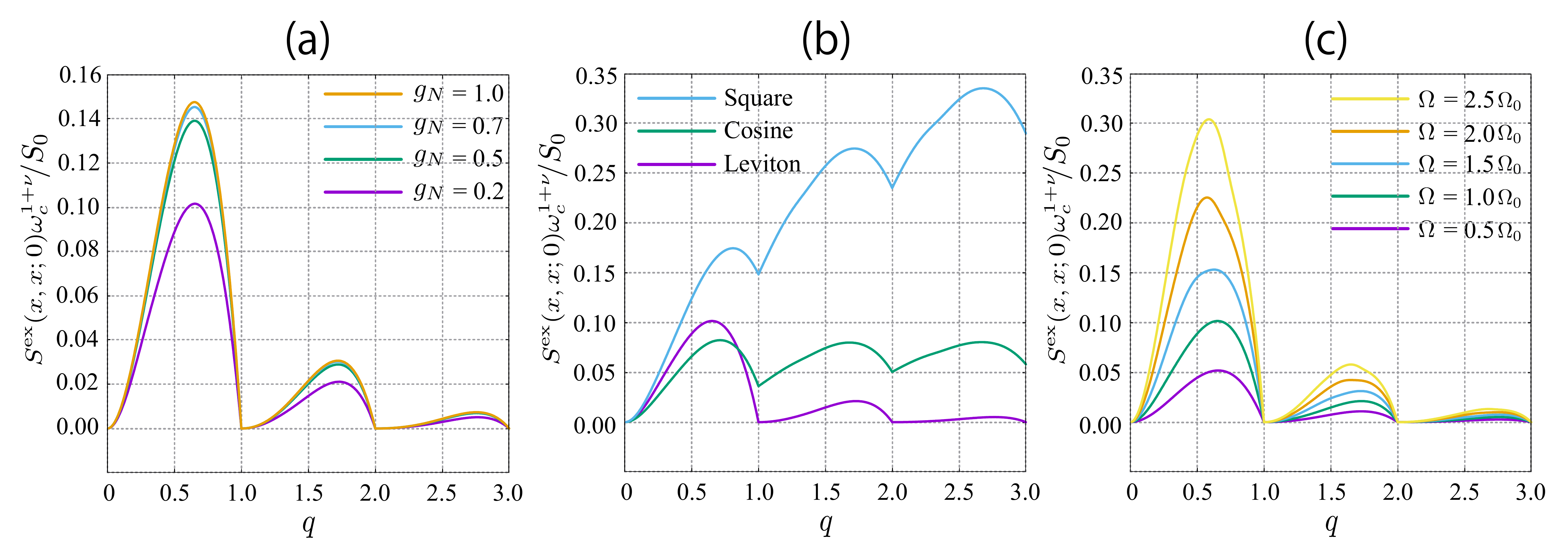}
    \caption{Excess noises as a function of $q$ for the two-lead case. 
    The unit of the noise is given by $S_0=4e^2\Gamma^2 /\pi^2 a^2$.
    (a) Leviton pulses for $\Omega = \Omega_0 \equiv \pi v_{\rm N}/L_2$ and $g_N = 1.0, 0.7, 0.5$, and $0.2$.
    (b) Three types of pulses for $\Omega = \Omega_0$ and $g_N=0.2$.
    (c) Leviton pulses for $g_N=0.2$ and $\Omega/\Omega_0 = 2.5, 2.0, 1.5, 1.0$, and $0.5$.
    In all the plots, we set $L_1=L/3$, $L_2=2L/3$,  $\eta=0.1$, and $\omega_c=100\Omega_0$.}
    \label{fig:2leads}
\end{figure*}

Let us first discuss the two-lead setup shown in Fig.~\ref{fig:Setup}~(a).
In Fig.~\ref{fig:2leads}~(a), we show the excess noise as a function of $q$
when Lorentzian voltage pulses (i.e. Levitons) are applied.
The interaction parameter of the CNT is set as $g_{\rm N}=1,0.7,0.5$ and $0.2$ and the other parameters are given in the caption.
The most striking feature is the fact that the excess noise is zero when $q$ is an integer, independently of the value of the interaction parameter. 
This is a well-known property of a Leviton drive for a non-interacting electronic
system.\cite{levitov96,keeling06,dubois13b}
Fig.~\ref{fig:2leads}~(a) shows that this property
remains valid even for strongly interacting non-chiral 1d electronic systems.
An analytical proof of this remarkable property is given in Sec.~\ref{sec:ZeroNoiseProof}.
For non-integer values of $q$, we observe that the excess noise is reduced a little as the Luttinger parameter $g_{\rm N}$ is reduced from unity.

Fig.~\ref{fig:2leads}~(b) shows the pulse-shape dependence of the excess noise, for $g_{\rm N}=0.2$. 
The excess noise vanishes only for Lorentzian pulses, while it does not for the cosine and square pulses.
These observations confirm that the exceptional features of the Levitons compared to all other voltage drives survive even for this interacting electron system.
We note that, contrarily to the case of quasi-particle tunneling into edge states of the fractional quantum Hall effect~\cite{rech17}, the excess noise does not show a singular behavior near integer values of $q$.
This is because we are considering here injection of electrons (rather than fractional quasi-particles) from the STM tip.

Fig.~\ref{fig:2leads}~(c) shows the excess noise for $g_{\rm N}=0.2$ as a function of $q$ for $\Omega/\Omega_0 = 0.5, 1.0, 1.5, 2.0$ and 2.5 and $\omega_c/\Omega_0 = 100$, where $\Omega_0$ is the unit of frequency.
Although the excess noise for a non-integer value of $q$ is almost proportional to $\Omega$, it grows a little faster than expected from linear dependence.

\subsection{One-lead case}
\label{sec:OneLeadCase}

\begin{figure*}
    \centering
    \includegraphics[width=18.cm]{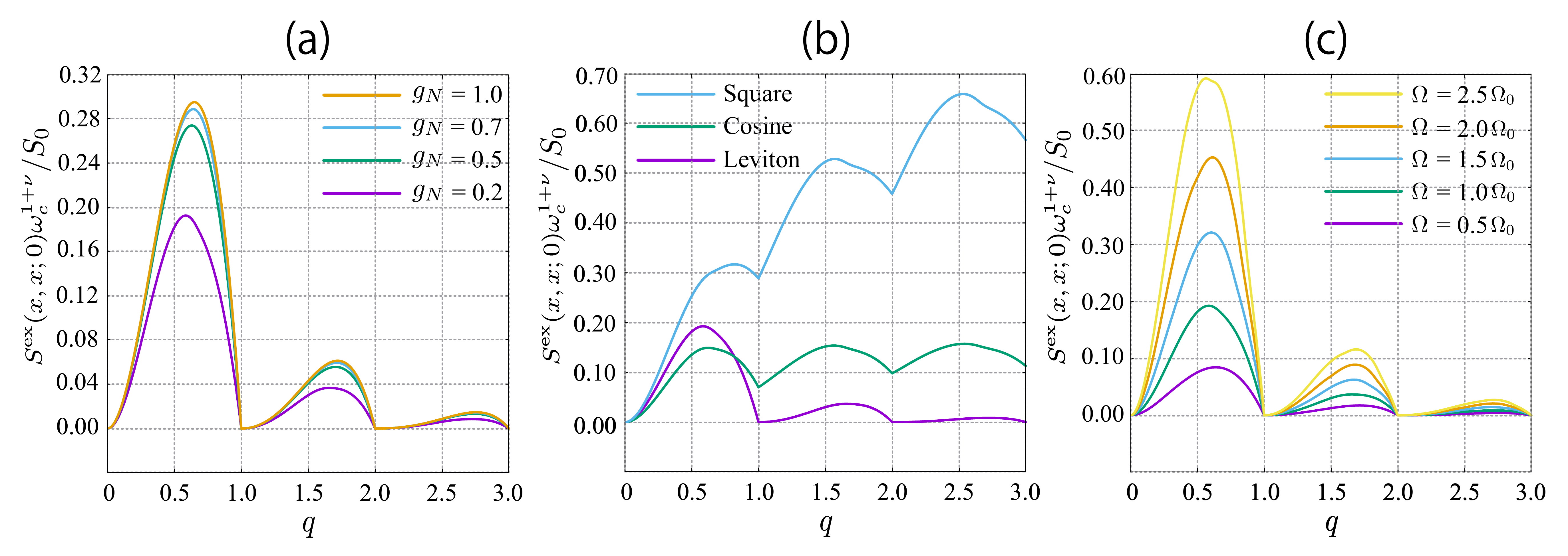}
    \caption{
    Excess noises as a function of $q$ for the one-lead case. 
    The unit of the noise is given by $S_0=4e^2\Gamma^2 /\pi^2 a^2$.
    (a) Leviton pulses for $\Omega = \Omega_0\equiv \pi v_{\rm N}/L_2$ and $g_N = 1.0, 0.7, 0.5$, and $0.2$.
    (b) Three types of pulses for $\Omega = \Omega_0$ and $g_N=0.2$.
    (c) Leviton pulses for $g_N=0.2$ and $\Omega/\Omega_0 = 2.5, 2.0, 1.5, 1.0$, and $0.5$.
    In all the plots, we set $L_1=L/3$, $L_2=2L/3$,  $\eta=0.1$, and $\omega_c=100\Omega_0$.
    }
    \label{fig:1lead}
\end{figure*}

Next, we discuss the one-lead setup shown in Fig.~\ref{fig:Setup}~(b).
We show the excess noise as a function of $q$ in Fig.~\ref{fig:1lead}~(a) for $g_{\rm N}=1, 0.7, 0.5$ and 0.2, where the other parameters are the same as Fig.~\ref{fig:2leads}.
We find that the qualitative features of the excess noise are similar to the two-lead case; for non-integer $q$ the excess noise is reduced a little as the interaction parameter $g_{\rm N}$ is reduced, while it vanishes when $q$ is an integer.
This means that the exceptional features of the Levitons are also
 valid for the interacting electrons systems for this setup.
Fig.~\ref{fig:1lead}~(b) and (c) show the pulse-shape and ac-frequency dependence, respectively.
By comparing with Fig.~\ref{fig:2leads}~(b) and (c), we find that the qualitative features are common with the two-lead case.
Compared to the two-lead case, the one-lead setup produces only small quantitative changes in the excess noise for non-integer values of $q$.

\subsection{Interference effects}
\label{sec:InterferenceEffect}

\begin{figure}
    \centering
    \includegraphics[width=6.cm]{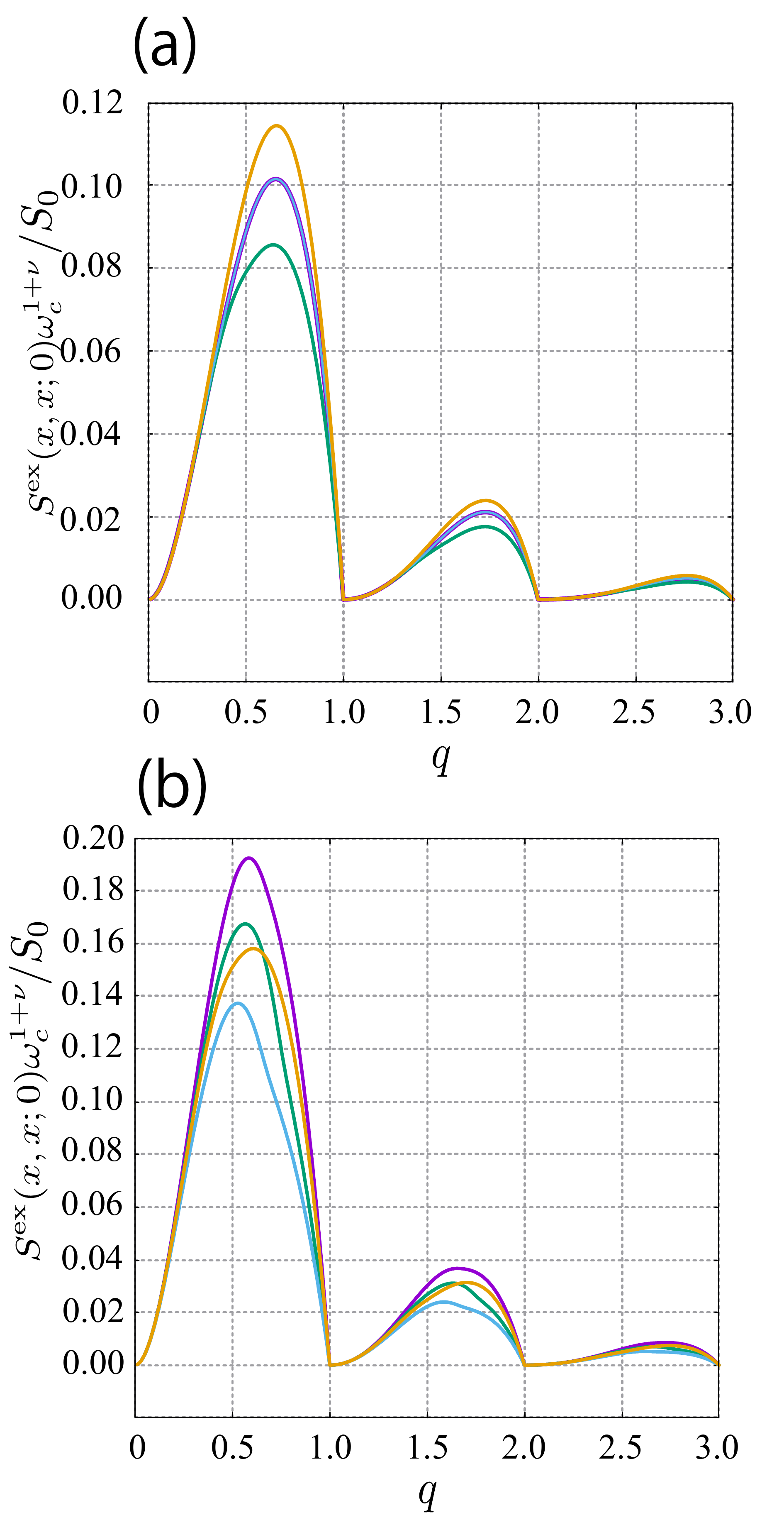}
    \caption{Geometry dependence of the excess noise for 
    (a) the two-lead case and (b) the one-lead case.
    We set the parameters as $(L_1,\Omega)=(L/3,3\pi v_{\rm N}/2L)$ (purple line), $(L/2,2\pi v_{\rm 
    N}/L)$ (green line), $(2L/3,3\pi v_{\rm N}/2L)$ (blue line), and $(L/2,\pi v_{\rm N}/L)$ (orange line).
    The other parameters are set as $\eta = 0.1$, $\omega_c/\Omega = 100$, and $g_{\rm N}=0.2$.}
    \label{fig:Ldep}
\end{figure}

As will be discussed in Sec.~\ref{sec:CurrentProfiles}, the Leviton pulses reflect at the boundary between the CNT and the leads in a complex manner and can interfere with subsequent Leviton pulses injected from the STM tip.
To see this interference effect, we show the excess noise in Fig.~\ref{fig:Ldep} for four geometries, i.e., four sets of the STM-tip position and two boundaries of the CNT.
Fig.~\ref{fig:Ldep}~(a) and (b) shows the two- and one-lead cases, respectively.
Although the geometry of the STM-tip and CNT boundaries controls the interference between Lorentzian pulses, the excess noise keeps its main features; it always vanishes when $q$ is an integer.
We find that the interference effect causes small changes in the excess noise for non-integer $q$ in both the two- and one-lead cases.
We conclude that the interference between the pulses only has a minor effect on the excess noise.

\subsection{Proof that excess noise is zero for integer Leviton}
\label{sec:ZeroNoiseProof}

As observed in Figs.~\ref{fig:2leads}, \ref{fig:1lead} and
\ref{fig:Ldep}, the excess noise is always zero
for a periodic drive of Levitons when $q$ is integer,
independently of the other parameters of the system 
(number of leads, position of the STM tip, frequency of the drive, etc.). We give here an analytical proof of
this property.

The excess noise is defined
as $S^{\rm ex}(x)=S(x) - e \bar{I}(x)$ in Eq.~(\ref{eq:Sexx}).
 Using $D(t)=D(-t)$ and $F(t)=-F(-t)$, we can recast Eq.~(\ref{eq:Sexx}) as:
\begin{align}
S^{\rm ex}(x)
&=\frac{4e^2 \Gamma^2}{N_{\rm lead}\pi^2 a^2} 
\Re \sum_{l=-\infty}^{\infty} \left| p_{l} \right|^2 
\int_{-\infty}^\infty \!\!\!dt \, \frac{D(t)}{(1+(v_Ft/a)^2)^{\frac{1}{2}}}  \nonumber \\ 
& \times \mbox{exp}\left[ i((q+l) \Omega t+ F(t)+\mathrm{ArcTan}(v_Ft/a))\right] ,
\label{eq:Sexxproof1}
\end{align}
where $F(t)$ and $D(t)$ are given by Eqs.~(\ref{eq:Ft})-(\ref{eq:Dt}), and $\Re$ denotes the real part.
$F(t)$ is composed by an infinite sum of terms, and each term has a corresponding
term in the infinite product which composes $D(t)$. A typical term in $F(t)$ can be written as:
\begin{equation}
    A \times \mbox{ArcTan}(\alpha(t)/\beta(t))
\end{equation}
and the corresponding term in $D(t)$ is:
\begin{equation}
    \left[  (\beta(t)/\gamma)^2 + (\alpha(t)/\gamma)^2\right]^{-A/2},
\end{equation}
with $\alpha(t) = 2 a v_F t$, $\beta(t)=a^2+(2kLg_{\rm N})^2-(v_Ft)^2$ 
(or a similar expression with $2 kL$ replaced by $2 kL + 2L_1$ or $2 k_L + 2 L_2$), and $A>0$, $\gamma>0$ are time-independent constants.
In the integral of Eq.~(\ref{eq:Sexxproof1}), this combination of terms appears as:
\begin{equation}
    \frac{\mbox{exp}(i A \times \mbox{ArcTan}(\alpha(t)/\beta(t)) )}
    {\left[  (\beta(t)/\gamma)^2 + (\alpha(t)/\gamma)^2\right]^ {A/2}}
   = \frac{\gamma^A}{(\beta(t) - i \alpha(t))^A}.
   \label{eq:Sexxproof2}
\end{equation}
From the expressions of $\alpha(t)$ and $\beta(t)$, one can check that the expression of Eq.~(\ref{eq:Sexxproof2}) has poles in the $t$ plane at:
\begin{equation}
    t= -\frac{i a}{v_F}  \pm \frac{2kLg_{\rm N}}{v_F} .
    \label{eq:Sexxproof3}
\end{equation}
Importantly, these two poles are in the lower half-plane.
The integrand of Eq.~(\ref{eq:Sexxproof1}) thus contains an infinite product of terms like Eq.~(\ref{eq:Sexxproof2}), which have poles and branch cuts in the lower half-plane only.
This product is multiplied by $\mbox{exp}(i (q+l) \Omega t)$; this factor allows to perform the integration by closing the contour in the upper (lower) complex plane for $q+l >0$ ($q+l<0$). 
As a Leviton drive for an integer charge $q$ is characterized 
by\cite{levitov96,ivanov97,keeling06,dubois13b,rech17}
\begin{equation}
    p_l =0, \quad \mbox{ for } \quad l \leq -q ,
\end{equation}
one can see that only terms with $q+l>0$ exist for a Leviton drive with integer charge $q$, which leads to a zero integral for the excess noise as the integrand has no poles in the upper complex plane.
The excess noise for a Leviton drive with an integer charge $q$ is thus always zero, independently of the value of the other parameters, including the interaction parameter $g_N$ in the nanotube.
We note that suppression of the excess noise has recently been discussed for a different setup using the non-chiral Luttinger liquid~\cite{Acciai2019}.

\section{Current profiles}
\label{sec:CurrentProfiles}

\begin{figure*}
    \centering
    \includegraphics[width=18cm]{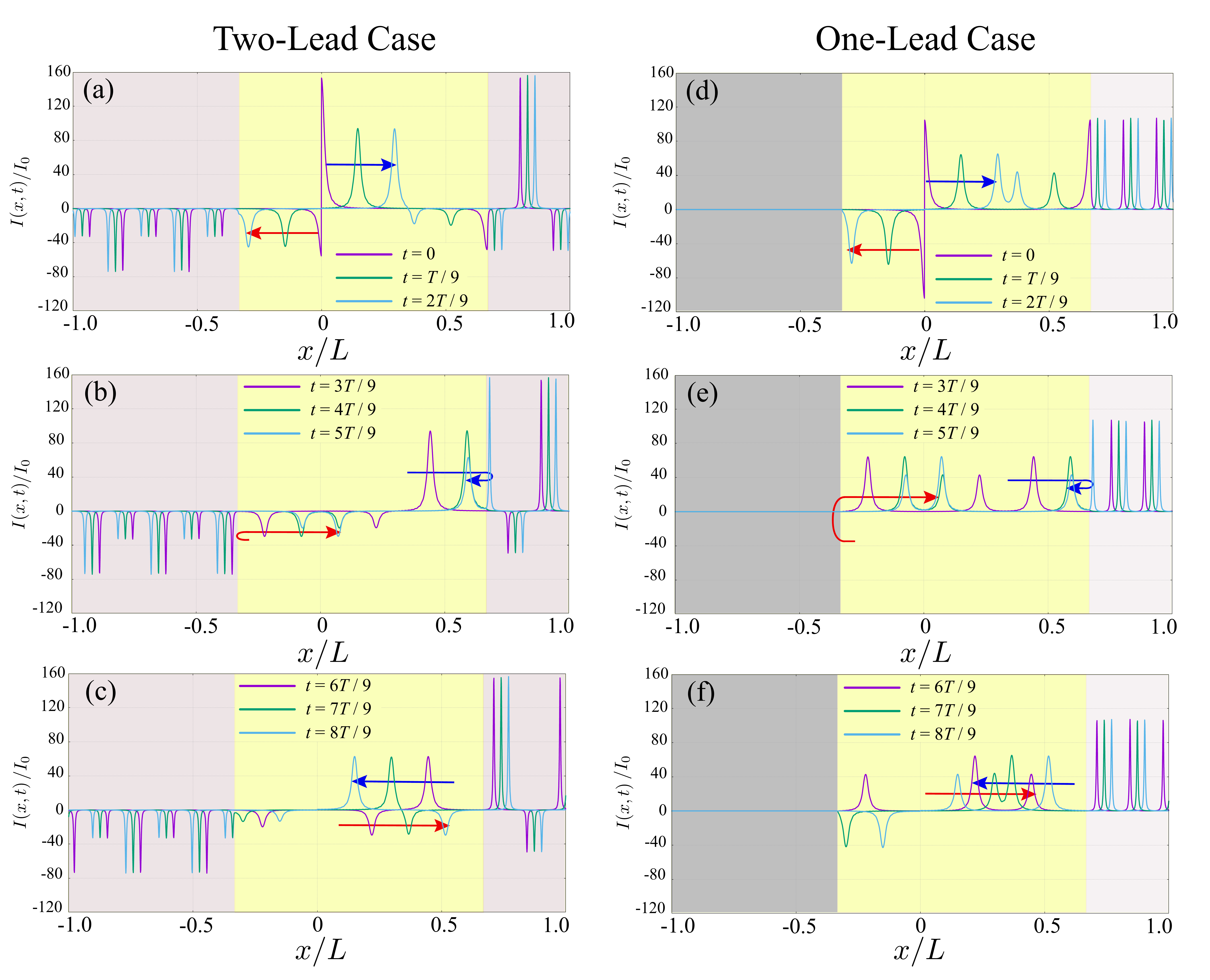}
    \caption{Time-dependent current profile as a function of position, inside the CNT (yellow background) and inside the normal leads (light gray background), for the two-lead case (left panels) and the one-lead case (right panels) in the case of $L_1=L/3$. Each panel shows the current at three different times as indicated. 
    The unit of the current is given by  $I_0=4e\Gamma^2/\pi^2 a^2$.
    The period of the ac voltage is given 
    $T=4L_1/v_{\rm N}=2L_2/v_{\rm N}$.
    Other experimental parameters are taken as $\eta=0.01$, $\omega_cT/2\pi=100$, and $g_{\rm N}=0.2$.}
     \label{fig:CurrentProfile}
\end{figure*}

Indeed, as the figures for the excess noise are qualitatively similar to the ones for a non-interacting system, one may think that the system behaves overall as a non-interacting one.
However, this is not the case, and it will be clear from the current profiles which are fundamentally different from the case of a non-interacting system.
It is thus quite remarkable that the excess noise is robust with respect to interactions, and in particular that it goes to zero for integer value of $q$ for a Leviton drive.

An essential element to understand the transport in the system is the reflection at each boundary between the nanotube and the leads.
These are characterized by the reflection coefficients $b_{1,2} = (g_N - g_{1,2})/(g_N+ g_{1,2})$. 
For the two-lead case, we have $b_1 = b_2 = (g_N-1)/(g_N+1) <0$, which means that an `electron'-like excitation is converted into a `hole'-like excitation by the reflection. This peculiar reflection, which is inherent to one-dimensional interacting electron systems, is called
Andreev reflection in analogy with that of a superconductor/normal junction\cite{safi95,Crepieux2003}.
For the one-lead case, one has $b_1=1$; the reflection at the right boundary is still an Andreev reflection, while at the left boundary it is a simple reflection.
The current profile will be the result of the interference of the injected pulses and their multiple reflections at the boundaries.

We first consider the two-lead case.
To visualize the Andreev reflection and interference with different Leviton pulses, we show the time-dependent current profile for $L_1 =L/3$ and $L_2 = 2L/3$ in Fig.~\ref{fig:CurrentProfile}~(a)-(c).
The expressions used to compute the time-dependent current profile
are detailled in Appendix~\ref{app:CurrentProfile}.
The period of the ac voltage is taken as $T=4L_1/v_{\rm N}=2L_2/v_{\rm N}$, where $v_{\rm N}=v_F/g_{\rm N}$ is the charge velocity in the CNT.
The three curves correspond to $t=0, T/9, 2T/9$ in (a), to $t=3T/9, 4T/9, 5T/9$ in (b), and to $t=6T/9, 7T/9, 8T/9$ in (c), respectively.
Fig.~\ref{fig:CurrentProfile}~(a) shows that the applied voltage pulse at $t=0$ creates two pulses propagating in opposite directions from the STM tip at $x=0$, as indicated by the blue and red arrows.
The heights of these two pulses are different
because of the resonance condition  $T=4L_1/v_{\rm N}=2L_2/v_{\rm N}$,
which means that pulses reflected one or several times at the left and right boundaries can interfere with new pulses created periodically at the tip position.
In Fig.~\ref{fig:CurrentProfile}~(b), the pulses are divided into a reflected part (in the CNT) and a transmitted part (in the leads) at the two boundaries of the CNT.
We note that the current carried by the reflected pulse has the same sign as the incident pulse despite the reversal of the propagation direction.
This is because the electron pulses are converted into hole pulses at the boundary due to the `Andreev' reflection characteristic of the Luttinger liquid~\cite{Crepieux2003,Guigou2007}.
In Fig.~\ref{fig:CurrentProfile}~(c), the reflected pulses continue to propagate in the CNT.
Here, because the lengths, $L_1$ and $L_2$, between the STM tip and the leads 1 or 2 are different, the timing of arrival of pulses at the origin is also different for the two reflected pulses.
For our choice of parameters, the pulse reflected at $x=L_2$ returns at the origin precisely after the period $T$ (see the blue arrow).
The fact that there is a strong interference between the pulse created at the STM tip and the pulses reflected on the right lead, while there is no such interference for the pulse
reflected at the left lead, explains the different current profile in the two metallic leads; the time-dependent current  is alternatively positive and negative in the right lead, while it is always negative in the left lead.

The time-dependent current profile for the one-lead case is shown in Fig.~\ref{fig:CurrentProfile}~(d)-(f) for the same geometry.
In contrast with the two-lead case, the system has one open boundary at $x=-L_1$, where the reflection is a standard one which keeps the sign of the charge excitation unchanged.
Therefore, the current carried by the pulse changes its sign after reflection at $x=-L_1$ (see Fig.~\ref{fig:CurrentProfile}~(e)) as in a usual reflection.
This change in the reflection properties affects the sign of the current in the right lead; the current is always positive there in contrast with the two-lead case (see Fig.~\ref{fig:CurrentProfile}~(a)-(c)).

\begin{figure*}
    \centering
    \includegraphics[width=18cm]{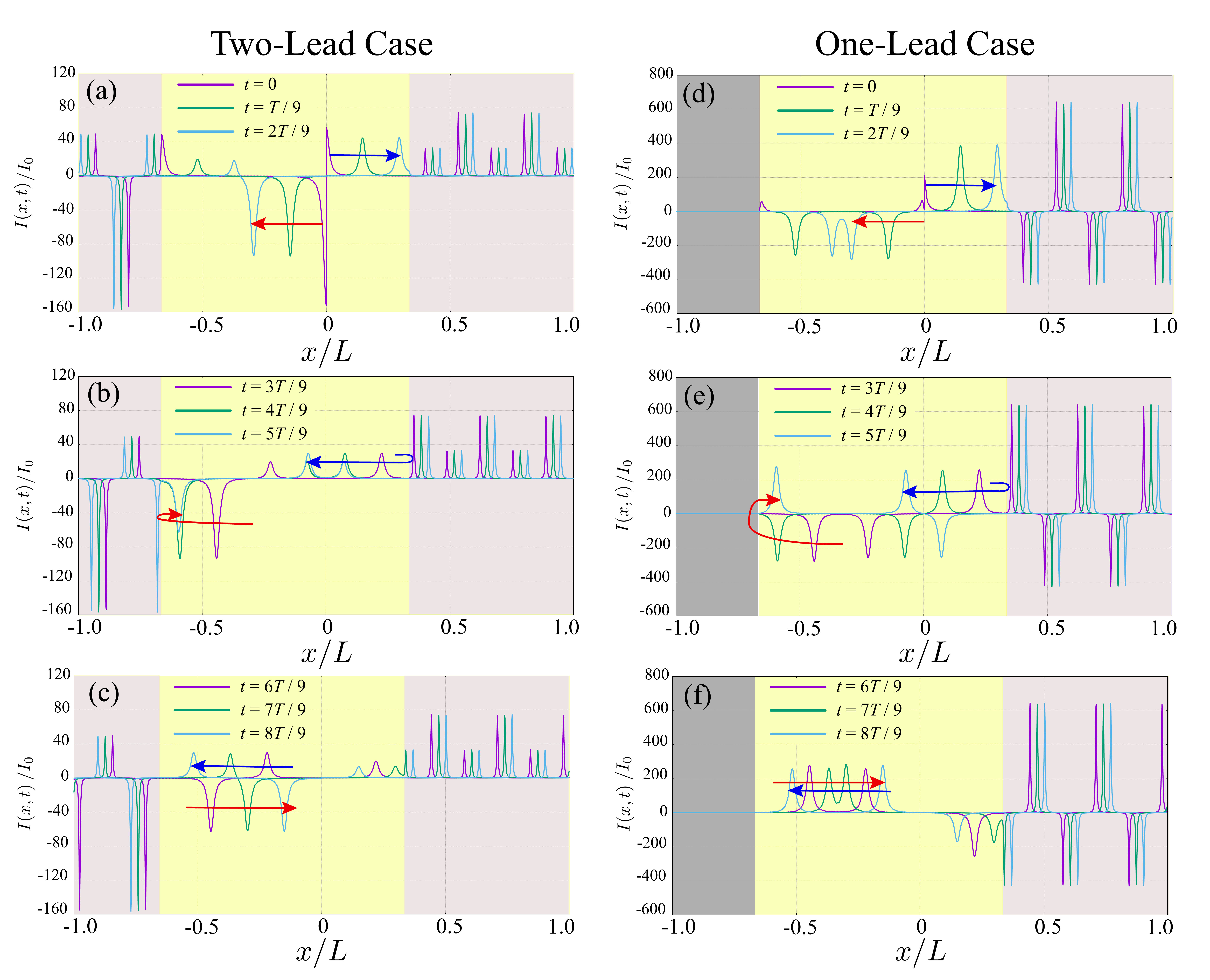}
    \caption{
    Time-dependent current profile as a function of position, inside the CNT (yellow background) and inside the normal leads (light gray background), for the two-lead case (left panels) and the one-lead case (right panels) in the case of $L_1=2L/3$. Each panel shows the current at three different times as indicated. 
    The unit of the current is given by  $I_0=4e\Gamma^2/\pi^2 a^2$.
    The period of the ac voltage is given 
    $T=2L_1/v_{\rm N}=4L_2/v_{\rm N}$.
    Other experimental parameters are taken as $\eta=0.01$, $\omega_cT/2\pi=100$, and $g_{\rm N}=0.2$.}
    \label{fig:CurrentProfile2}
\end{figure*}

We further show the current profile for a different geometry with $L_1 =2L/3$ and $L_2 = L/3$ in Fig.~\ref{fig:CurrentProfile2}.
Although the reflection properties at the boundaries of the CNT are the same as the previous case, the geometry affects the current profile in the metallic leads due to the modifications of the interference conditions between pulses.
For the two-lead case (Fig.~\ref{fig:CurrentProfile2}~(a)-(c)), the current is now always positive in the right lead, while it can be both negative and positive in the left one.
This result is the inverse of what was observed with the previous geometry (Fig.~\ref{fig:CurrentProfile}~(a)-(c)).
Also for the one-lead case (Fig.~\ref{fig:CurrentProfile2}~(d)-(f)), the current can now be negative in the right lead, in contrast with the previous geometry (Fig.~\ref{fig:CurrentProfile}~(d)-(f)).

\section{Summary}
\label{sec:Summary}

We theoretically studied the effect of electron correlations in single electron injection into a carbon nanotube (CNT) coupled with metallic leads.
We formulated the current and the excess noise induced by an AC voltage driving by describing the CNT in terms of a Luttinger liquid, and studied how electron correlations modified (or not) the property of minimal excess noise for Lorenzian pulse (Leviton) injection.
We showed both analytically and numerically that the excess noise vanishes when each Leviton pulse includes an integer number of electrons, as observed in non-interacting electron systems.
This indicates that the electron correlations in the CNT do not change the physics of minimum noise pulses, i.e. Levitons, at all.
For a non-integer electron injection, the excess noise depends on both the geometry of the system and the ac driving frequency.
We also showed that the time-dependent current profile depends on the geometry and the ac frequency through the interference condition between injected pulses.
We demonstrated that injected pulses induces Andreev-like (resp. normal) scattering at the junction with a metallic lead (resp. at the open boundary).
This affects the excess noise in non-integer electron injection.
Our findings show the universality of the minimum noise properties for the Leviton pulses, which holds even in interacting electron systems. 
Detailed setup and estimate for experimental verification of our results are left as a future problem.
Our calculations could also apply to artificially created non-chiral Luttinger liquid systems in the quantum Hall
effect.\cite{brasseur2017}

\begin{acknowledgments} 
This French-Japanese collaboration is supported by the CNRS International Research Project ``Excitations in Correlated Electron Systems driven in the GigaHertz range" (ESEC). This work received support from the French government under the France 2030 investment plan, as part of the Initiative d'Excellence d'Aix-Marseille Universit\'e - A*MIDEX, through the institutes IPhU (AMX-19-IET-008) and AMUtech (AMX-19-IET-01X).
T. K. acknowledges support from the Japan Society for the Promotion of Science (JSPS KAKENHI Grant No.~JP20K03831). 
\end{acknowledgments}

\appendix

\section{Green functions}
\label{app:Green}

\

In this appendix, we show the main steps
of the calculation to obtain the Green function of the bosonic fields for an inhomogeneous infinite one-dimensional system, with an interaction
parameter being a piecewise constant function defining three different regions, without any asumption of spatial symmetry.\cite{Guigou2007}

Let us first consider the imaginary-time Green functions defined by
\begin{align}
G^{XY}_{j\delta}(x,x',\tau)&=
\left\langle X_{j\delta}(x,\tau) Y_{j\delta}(x',0) \right\rangle ,
\label{eq:GreenDef}
\end{align}
for $0 < \tau < \hbar \beta$, where $X,Y=\theta$ or $\phi$ and $X_{j\delta}(x,\tau) = e^{H\tau/\hbar} X_{j\delta}(x) e^{-H\tau/\hbar}$ indicates the imaginary-time evolution.
From the Hamiltonian (\ref{HamCNT}), the Green functions,  $G^{\phi\phi}_{j\delta}(x,x',\tau)$ and $ G^{\theta\theta}_{j\delta}(x,x',\tau)$, obey the following equations:
\begin{align}
& -\left(\frac{g_{j\delta}(x)}{v_{j\delta}(x)}\partial_\tau^2+\partial_x v_{j\delta}(x) g_{j\delta}(x) \partial_x \right) G^{\phi\phi}_{j\delta}(x,x',\tau) \nonumber \\
&\hspace{45mm} =\delta(x-x') \delta(\tau), 
\label{eq:EOMphiphi} \\
&-\left(\frac{\partial_\tau^2}{v_{j\delta}(x) g_{j\delta}(x)}+\partial_x \frac{v_{j\delta}(x)}{g_{j\delta}(x)}\partial_x \right) G^{\theta \theta}_{j\delta}(x,x',\tau) \nonumber\\
&\hspace{45mm} =\delta(x-x') \delta(\tau), \label{eq:EOMthetatheta}
\end{align}
where $\delta(x)$ is a delta function.
The mixed Green functions, $G^{\phi \theta}_{j\delta}$
and $G^{\theta \phi}_{j\delta}$, can be obtained from $G^{\phi\phi}_{j\delta}$ and $G^{\theta\theta}_{j\delta}$ as
\begin{align}
i\partial_{\tau} G_{j\delta}^{\phi\theta}(x,x',\tau) &= \frac{v(x)}{g(x)} \partial_x G_{j\delta}^{\theta\theta}(x,x',\tau) , \label{eq:Gphitheta}\\
i\partial_\tau G_{j\delta}^{\theta\phi}(x,x',\tau) &= v(x)g(x) \partial_x G_{j\delta}^{\phi\phi}(x,x',\tau) . \label{eq:Gthetaphi}
\end{align}
Using the fact that the interaction parameters $g_{j\delta}(x)$ and the velocities $v_{j\delta}(x)$ are piecewise constant functions, with three different domains: 
$x<-L_1$ (left lead), $-L_1<x<L_2$ (nanotube) and $x>L_2$ (right lead), one can solve for the Green functions $G^{\phi \phi}_{j\delta}$ and $G^{\theta \theta}_{j\delta}$ in the Matsubara frequency space.
For calculation of the current and current noise, the Green functions at $x=x'=0$ (which is the injection point) and its spatial derivatives in which $x$ or $x'$ is set as zero are required.

Since the full derivation is rather lengthy, we only explain here the calculation of $G_{c+}^{\phi\phi}(x,x',\tau)$.
The other types of the Green functions can be calculated in a similar way.
By the Fourier transformation of Eq.~(\ref{eq:EOMphiphi}), we obtain
\begin{align}
& \left( \frac{g_{c+}(x)}{v_{c+}(x)}\omega^2 -\partial_x v_{c+}(x)g_{c+}(x)\partial_x\right)G_{c+}^{\phi\phi}(x,x',\omega) \nonumber \\
& \hspace{30mm} =\delta(x-x'),
\end{align}
where $\omega$ is a Matsubara frequency, that can be regarded as a real number at zero temperature, and 
\begin{align}
G^{\phi\phi}_{c+}(x,x',\omega) =\int d\tau \, G_{c+}^{\phi\phi}(x,x',\tau)e^{i\omega\tau} .
\end{align}
It is straightforward to solve this differential equation with respect to $x$.
For example, the Green function in the range of $-L_1<x'<L_2$ is obtained as
\begin{align}
&G^{\phi\phi}_{c+}(x,x',\omega) \nonumber \\
&= \left\{ \begin{array}{ll}
A(x')e^{\frac{|\omega|x}{v_1}}, & (x<-L_1), \\
B(x')e^{\frac{|\omega|x}{v_{\rm N}}}+C(x')e^{-\frac{|\omega|x}{v_{\rm N}}}, & (-L_1<x<x'), \\
D(x')e^{\frac{|\omega|x}{v_{\rm N}}}+E(x')e^{-\frac{|\omega|x}{v_{\rm N}}}, & (x'<x<L_2), \\
F(x')e^{-\frac{|\omega|_n x}{v_2}}, & (L_2<x),
\end{array} \right.
\end{align}
where
\begin{align}
A(x')&=\frac{2g_{\rm N}}{g_1+g_{\rm N}} e^{\frac{|\omega|L_1}{v_1}-\frac{|\omega|L_1}{v_{\rm N}}} B(x'), \\
B(x')&=\frac{1}{2g_{\rm N}|\omega|}
\frac{  e^{-\frac{|\omega|}{v_N}x'}+\gamma_2 e^{\frac{|\omega|}{v_N}x'}}{1-\gamma_1 \gamma_2},  \\
C(x')&=\gamma_1 B(x'), \\
D(x')&=\gamma_2 E(x'),\\
E(x')&=\frac{1}{2g_{\rm N}|\omega|}
\frac{  e^{\frac{|\omega|}{v_{\rm N}}x'}+\gamma_1 e^{-\frac{|\omega|}{v_{\rm N}}x'}}{1-\gamma_1 \gamma_2}, \\
F(x')&=\frac{2g_{\rm N}}{g_2+g_{\rm N}} e^{\frac{|\omega|L_2}{v_2}-\frac{|\omega|L_2}{v_{\rm N}}}E(x').
\end{align}
Here, the coefficients, $\gamma_1$ and $\gamma_2$, are given as
\begin{align}
\gamma_1 &= \frac{g_{\rm N}-g_1}{g_{\rm N}+g_1} e^{\frac{-2|\omega|L_1}{v_{\rm N}}}
\equiv b_1 e^{-\frac{2|\omega|L_1}{v_{\rm N}}},\\
\gamma_2 &= \frac{g_{\rm N}-g_2}{g_{\rm N}+g_2} e^{\frac{-2|\omega|L_2}{v_{\rm N}}}
\equiv b_2 e^{\frac{-2|\omega|L_2}{v_{\rm N}}}.
\end{align}
By setting $x,x'\rightarrow 0$, we obtain
\begin{align}
G_{c+}^{\phi\phi}(0,0,\omega) = \frac{1}{2|\omega|g_{\rm N}}
\frac{(1+\gamma_1)(1+\gamma_2)}{1-\gamma_1 \gamma_2} .
\end{align}
Here, we further use the expansion
\begin{align}
\frac{1}{1-\gamma_1\gamma_2}
= \sum_{k=1}^\infty (b_1 b_2)^k
e^{-\frac{2kL|\omega|}{v_{\rm N}}},
\end{align}
where $k$ can be regarded as the number of round trips of a pulse in the CNT.
The inverse Fourier transformation can easily be performed by using the formula
\begin{align}
\partial_\tau \int_0^\infty \frac{d\omega}{\omega} e^{\omega(x\pm i\tau)} = \frac{\mp i}{x\pm i\tau},
\end{align}
and by the analytic continuation $\tau = it + a/v_{\rm F}$ the Green function is calculated as
\begin{align}
& \tilde{G}_{c+}^{\phi\phi}(0,0,t)  \nonumber \\
&= -\frac{1}{2\pi g_{\rm N}} \Biggl\{ \log (1+iv_{\rm F}t/a) + \sum_{k=1}^{\infty} (b_1 b_2)^k I_+(2kL) \nonumber \\
& \hspace{10mm} + \frac{1}{2}
\sum_{k=0}^{\infty} b_1 (b_1 b_2)^k I_+(2kL+2L_1) \nonumber \\
& \hspace{10mm} +\frac{1}{2}
\sum_{k=0}^{\infty} b_2 (b_1 b_2)^k I_+(2kL+2L_2) \Biggr\}, 
\end{align}
where
\begin{align}
I_{\pm}(x) &= \pm \log \left( 1+ \frac{iv_{\rm F}t}{a+ix g_{\rm N}} \right) \nonumber \\
&+ \log \left( 1+ \frac{iv_{\rm F}t}{a-ixg_{\rm N}} \right) .
\end{align}
Here, the real-time Green's function is defined as
\begin{align}
\tilde{G}^{XY}_{j\delta}(x,x',t)&=
\left\langle X_{j\delta}(x,t) Y_{j\delta}(x',0) \right\rangle ,
\nonumber \\
& - \left \langle X_{j\delta}(x,t)^2 \right \rangle - \left \langle Y_{j\delta}(x',0)^2 \right \rangle,
\label{eq:GreenDef2}
\end{align}
where $X,Y=\theta$ or $\phi$ and $X_{j\delta}(x,t) = e^{iHt/\hbar} X_{j\delta}(x) e^{-iHt/\hbar}$ indicates the real-time evolution.
In a similar way, the other Green functions can be calculated from Eqs.~(\ref{eq:EOMthetatheta})-(\ref{eq:Gthetaphi}) as
\begin{align}
& \tilde{G}_{c+}^{\theta\theta}(0,0,t)  \nonumber \\
&= -\frac{g_{\rm N}}{2\pi} \Bigl\{ \log (1+iv_{\rm F}t/a) + \sum_{k=1}^{\infty} (b_1 b_2)^k I_+(2kL) \nonumber \\
& \hspace{10mm} -\frac{1}{2}
\sum_{k=0}^{\infty} b_1 (b_1 b_2)^k I_+(2kL+2L_1) \nonumber \\
& \hspace{10mm} -\frac{1}{2}
\sum_{k=0}^{\infty} b_2 (b_1 b_2)^k I_+(2kL+2L_2) \Bigr\}, \\
& \tilde{G}_{c+}^{\theta\phi}(0,0,t) = -\frac{1}{4\pi} \Bigl\{
\sum_{k=0}^\infty b_1 (b_1b_2)^k I_-(2kL+2L_1) \nonumber \\
& \hspace{10mm} - \sum_{k=0}^\infty b_2(b_1b_2)^k I_-(2kL+2L_2) \Bigr\}, 
\end{align}
\begin{align}
& \tilde{G}_{c+}^{\phi\theta}(0,0,t) = \frac{1}{4\pi} \Bigl\{
\sum_{k=0}^\infty b_1 (b_1b_2)^k I_-(2kL+2L_1) \nonumber \\
& \hspace{10mm} - \sum_{k=0}^\infty b_2(b_1b_2)^k I_-(2kL+2L_2) \Bigr\}, 
\end{align}
For other modes, i.e., $(j,\delta) = (c,-), (s,+), (s,-)$, the Green functions are easily obtained from the above results by setting $g_1=g_2=g_{\rm N}=1$ (the two-lead case) or $g_1=0$ and $g_2=g_{\rm N}=1$ (the one-lead case).

Next, we calculate spatial derivatives of the Green functions for the channel $(j,\delta)=(c,+)$.
For simplicity, we drop the subscripts assigning the mode hereafter.
By solving the Fourier transformation of Eqs.~(\ref{eq:EOMphiphi})-(\ref{eq:Gthetaphi}), we obtain
\begin{align}
& G^{\phi\phi}(x,0,\omega) = G^{\phi\phi}(0,x,\omega) = \frac{1}{g_{\rm N}+g_2} \frac{1}{|\omega|}G_+(x), \\
& G^{\theta\theta}(x,0,\omega) = G^{\theta\theta}(0,x,\omega) = 
\frac{g_{\rm N}g_2}{g_{\rm N}+g_2} \frac{1}{|\omega|}G_-(x), \\
& G^{\phi\theta}(x,0,\omega) = 
\frac{g_{\rm N}}{g_{\rm N}+g_2} \frac{1}{\omega}G_-(x), \\
& G^{\phi\theta}(0,x,\omega) = 
-\frac{g_2}{g_{\rm N}+g_2} \frac{1}{\omega}G_+(x), \\
& G^{\theta\phi}(x,0,\omega) = 
\frac{g_2}{g_{\rm N}+g_2} \frac{1}{\omega}G_+(x),\\
& G^{\theta\phi}(0,x,\omega) = 
-\frac{g_{\rm N}}{g_{\rm N}+g_2} \frac{1}{\omega}G_-(x), \\
& G_{\pm}(x) = \frac{1\pm \gamma_1}{1-\gamma_1\gamma_2}
e^{-b|\omega|/v_{\rm N}-(x-b)|\omega|/v_2} ,
\end{align}
for $x>L_2$.
By using the same techniques for calculation of $\partial_\tau G^{XY}(0,0,\tau)$, the spatial derivatives can be calculated.
For example, we obtain
\begin{align}
& \partial_x G^{\phi\phi}(x,0,t) = \partial_x G^{\phi\phi}(0,x,t) \nonumber \\
& = - \frac{1}{2\pi(g_{\rm N}+g_2)v_2} \nonumber \\
& \times \sum_{k=0}^\infty (b_1b_2)^k \Bigl[ 
\frac{1}{\alpha_k(x)+t-i\tau_0}
+ \frac{1}{\alpha_k(x)-t+i\tau_0} 
\nonumber \\
&\hspace{5mm} +   \frac{b_1}{\beta_k(x)+t-i\tau_0}
+ \frac{b_1}{\beta_k(x)-t+i\tau_0} 
\Bigr] ,\\
& \alpha_k(x) = \frac{2kL+L_2}{v_{\rm N}} + \frac{x-L_2}{v_2}, \\
& \beta_k(x) = \alpha_k(x) + \frac{2L_1}{v_{\rm N}},
\end{align}
after analytic continuation $\tau = it +\tau_0$, where $\tau_0 = a/v_{\rm F}$ is a short-time cutoff.

The Green functions obtained above can be related to the Keldysh Green functions defined as
\begin{align}
& G^{XY}_{K}(x,x',t) =\langle T_K \{ X(x,t^{\eta_1}) Y(x',0^{\eta_2}) \} \rangle ,
\end{align}
where $\eta_1$, $\eta_2$ ($=\pm 1$) represents the forward ($+1$) or backward ($-1$) contours.
The Keldysh Green functions are expressed in a matrix form as
\begin{align}
& G^{XY}_{K}(x,x',t) = \left( \begin{array}{cc}
G^{XY}_{(++)}(x,x',t) & G^{XY}_{(+-)}(x,x',t) \\
G^{XY}_{(-+)}(x,x',t) & G^{XY}_{(--)}(x,x',t) 
\end{array} \right) \nonumber \\
&= \left( \begin{array}{cc}
G^{XY}(x,x',|t|) & G^{XY}(x',x,-t) \\
G^{XY}(x,x',t) & G^{XY}(x',x,-|t|) 
\end{array} \right).
\end{align}
The current and current noise are calculated combining these expressions for $G^{XY}_{(\eta_1\eta_2)}(x,x',t)$ with the results given in Ref.~\onlinecite{Guigou2007}.

\newpage

The Keldysh Green functions of the bosonic field describing the STM tip is calculated as
\begin{align}
& g_{\sigma(\eta_1\eta_2)}(t_1-t_2) =\langle T_K \{ \tilde{\varphi}_\sigma (t_1^{\eta_1}) \tilde{\varphi}_\sigma (t_2^{\eta_2})\} \rangle \nonumber \\
&\hspace{5mm} =-\log \Bigl[1+i(\eta_1+\eta_2)\frac{v_F|t_1-t_2|}{2a}
\nonumber \\
&\hspace{15mm}- i(\eta_1-\eta_2)\frac{v_F(t_1-t_2)}{2a}
\Bigr].
\label{eq:STMGreen}
\end{align}

\section{Current Profile}
\label{app:CurrentProfile}

The calculation of $I(x,t)$ for arbitrary time and position is
a rather long, tedious, but straightforward extension of that in Refs.~\onlinecite{Crepieux2003,Guigou2007}.
We only show the final expression as follows:
\begin{widetext}
\begin{align}
I(x,t)&=\frac{-8iev_F \Gamma^2}{\pi^2 a^2}\sum_{l,l'} p_l p_{l'}^* e^{i(l-l')\Omega t}  \nonumber\\
&\qquad \times \left[ \frac{i \theta((l'-l)\Omega)}{(g_{\rm N}+g_{2})v_{2}} \left(\frac{1+b_1 e^{-i\frac{2(l-l')\Omega L_1}{v_N}}}{1-b_1b_2 e^{-i\frac{2(l-l')\Omega}{v_N}L}}e^{-i\frac{(l-l')\Omega L_2}{v_N}}e^{-i\frac{(l-l')\Omega}{v_2}(x-L_2)} -{\rm c.c.}\right) X_{1,(ll')} \right. \nonumber\\
&\qquad \qquad \left.-\frac{i}{(g_{\rm N}+g_2)v_2} \left(\frac{1+b_1 e^{-i\frac{2|(l-l')\Omega|L_1}{v_N}}}{1-b_1b_2 e^{-i\frac{2|(l-l')\Omega|}{v_N}L}}e^{-i\frac{|(l-l')\Omega| L_2}{v_N}}e^{-i\frac{|(l-l')\Omega|}{v_2}(x-L_2)} \right) X_{2,(ll')} \right],\qquad (x>L_2),\\
I(x,t)&=\frac{-8iev_F \Gamma^2}{\pi^2 a^2}\sum_{l,l'} p_l p_{l'}^* e^{i(l-l')\Omega t}  \nonumber\\
&\qquad \times \left[ \frac{i \theta((l'-l)\Omega)}{2v_F} 
\left(\frac{1+b_1 e^{-i\frac{2(l-l')\Omega L_1}{v_N}}}{1-b_1b_2 e^{-i\frac{2(l-l')\Omega}{v_N}L}}(e^{-i\frac{(l-l')\Omega}{v_N}x}-b_2 e^{i\frac{(l-l')\Omega}{v_N}(x-2L_2)})-{\rm c.c.}\right)X_{1,(ll')} \right.  \nonumber\\
&\qquad \qquad \left.-\frac{i}{2v_F} \left(\frac{1+b_1 e^{-i\frac{2|(l-l')\Omega|L_1}{v_N}}}{1-b_1b_2 e^{-i\frac{2|(l-l')\Omega|}{v_N}L}}(e^{-i\frac{|(l-l')\Omega|}{v_N}x}-b_2 e^{i\frac{|(l-l')\Omega|}{v_N}(x-2L_2)})\right) X_{2,(ll')}\right],\qquad (L_2\geq x \geq 0),
\end{align}
\begin{align}
&X_{1,(ll')}=\int_0^\infty d\tau D(\tau) \sin \left[ \left(\omega_0+\frac{l+l'}{2}\Omega\right)\tau \right]
\left(\frac{\sin\left(\frac{l-l'}{2}\Omega \tau\right) e^{-iF(\tau)}}{1+iv_F \tau/a}-{\rm Im}\left[\frac{e^{i(l-l')\Omega \tau/2}e^{-iF(\tau)}}{1+iv_F\tau/a} \right] \right), \\
&X_{2,(ll')}=\int_0^\infty d\tau D(\tau) \sin \left[ \left(\omega_0+\frac{l+l'}{2}\Omega\right)\tau \right]
\left(\frac{\sin\left(\frac{l-l'}{2}\Omega \tau\right) e^{iF(\tau)}}{1-iv_F \tau/a}+{\rm Im}\left[\frac{e^{-i(l-l')\Omega \tau/2}e^{-iF(\tau)}}{1+iv_F\tau/a} \right] \right).
\end{align}
\end{widetext}

\bibliography{reference}

\end{document}